\documentclass[sigconf]{acmart}

\usepackage{multirow}
\usepackage{fontawesome5}
\usepackage{subcaption}
\usepackage{ulem}

\newcommand{\rqi}{How does the merge rate of agentic-PRs change after the creation of instruction files?}
\newcommand{\rqii}{How does task complexity that agents undertake change after the creation of instruction files?}
\newcommand{\rqiii}{How does the efforts to merge agentic-PRs change after the creation of instruction files?}
\newcommand{\rqiv}{What is the relation between the verbosity of instruction files and the merge rates of agentic-PRs?}

\AtBeginDocument{%
  }

\copyrightyear{2026}
\acmYear{2026}
\setcopyright{cc}
\setcctype{by-nc-nd}
\acmConference[MSR '26]{23rd International Conference on Mining Software Repositories}{April 13--14, 2026}{Rio de Janeiro, Brazil}
\acmBooktitle{23rd International Conference on Mining Software Repositories (MSR '26), April 13--14, 2026, Rio de Janeiro, Brazil}
\acmPrice{}
\acmDOI{10.1145/3793302.3793601}
\acmISBN{979-8-4007-2474-9/2026/04}

\newboolean{showcomments}
\setboolean{showcomments}{true}

\ifthenelse{\boolean{showcomments}}
{\newcommand{\nbc}[3]{
 {\colorbox{#3}{\bfseries\sffamily\scriptsize\textcolor{white}{#1}}}
 {\textcolor{#3}{\sf\small$\blacktriangleright$\textit{#2}$\blacktriangleleft$}}
 }
}
{\newcommand{\nbc}[3]{}
}

\newcommand{\med}[1]{\nbc{Mohammed}{#1}{blue}}

\begin{document}

%%
%% The "title" command has an optional parameter,
%% allowing the author to define a "short title" to be used in page headers.
\title{Toward Instructions-as-Code: Understanding the Impact of Instruction Files on Agentic Pull Requests}

%%
%% The "author" command and its associated commands are used to define
%% the authors and their affiliations.
%% Of note is the shared affiliation of the first two authors, and the
%% "authornote" and "authornotemark" commands
%% used to denote shared contribution to the research.
\author{Ali Arabat, Mohammed Sayagh}
\affiliation{%
  \institution{École de Technologie Supérieure}
  \city{Montréal}
  \country{Canada}
}
\email{ali.arabat.1@ens.etsmtl.ca, mohammed.sayagh@etsmtl.ca}

%%
%% By default, the full list of authors will be used in the page
%% headers. Often, this list is too long, and will overlap
%% other information printed in the page headers. This command allows
%% the author to define a more concise list
%% of authors' names for this purpose.
\renewcommand{\shortauthors}{Arabat and Sayagh}

%%
%% The abstract is a short summary of the work to be presented in the
%% article.
\begin{abstract}
AI-agents (e.g., GitHub Copilot) collaborate as teammates in different software engineering tasks, including code generation proposed through pull requests (Agentic-PRs). For better agent efficiency, developers create instruction files that guide the AI-agents, including how to navigate the project, locate the right components, run tests, respect best practices, and more. In this paper, we investigate the relationship between the creation of these instructions and the performance of AI-agents in creating better pull requests, which have a higher chance of success (i.e., the merge rate), address more complex tasks (e.g., code churn), and require less effort to be merged (e.g., time to merge). To this end, we analyze 15,549 agentic PRs from 148 projects in the AIDev dataset. Using the three dimensions, we compare each project before and after the creation of the instruction files. We find that specifying instructions for AI-agents does not necessarily lead to better results. With the instruction files, 27.7\% of the projects increased their merge rate by at least 20\%, while 26.35\% decreased it. The same observation is seen with the amount of changes (e.g., code churn, number of modified files) and with the efforts to merge an agentic PR (e.g., merge time and number of comments). From a first exploration, we find that projects that managed to increase their merge rate have substantially longer instruction files, which are also well structured into a higher number of sections and sub-sections. Our results motivate the need for research to assist practitioners in framing the development of instruction files as a software engineering activity (aka, \textbf{Instructions-as-Code}).
\end{abstract}

%%
%% The code below is generated by the tool at http://dl.acm.org/ccs.cfm.
%% Please copy and paste the code instead of the example below.
%%
  

%%
%% Keywords. The author(s) should pick words that accurately describe
%% the work being presented. Separate the keywords with commas.
%% A "teaser" image appears between the author and affiliation
%% information and the body of the document, and typically spans the
%% page.

\received{25 December 2025}
\received[revised]{19 Junuary 2026}
\received[accepted]{19 January 2026}

%%
%% This command processes the author and affiliation and title
%% information and builds the first part of the formatted document.
\maketitle

\section{Introduction}

With LLMs, software engineering tasks have undergone unprecedented transformation~\cite{10.1145/3695988}, allowing Autonomous coding agents such as Claude~\cite{anthropic2024claude}, Cursor~\cite{cursor2024}, and Copilot~\cite{10.1016/j.jss.2023.111734} to be integrated into software development workflows. These agents support a wide range of activities, including code generation~\cite{10.1145/3597503.3639219}, code review~\cite{11121707}, and testing~\cite{10.1145/3597503.3639180}. These tools have become essential in developers' daily routines due to their ability %not only to collaborate with humans but also 
to perform complex, repository-wide tasks that extend beyond isolated code snippets. They are nowadays integrated as teammates to software processes~\cite{li2025riseaiteammatessoftware}.

To help agents better perform their tasks, developers write
custom instruction files, which provide agents with project-specific guidance and context~\cite{githubdocs_copilot_instructions}.
These instructions specified in files (e.g., .github/copilot-instructions.md) guide coding agents in understanding and operating within a repository, specifying essential procedures for building, testing, and validating changes, instructing the agents on the project-specific best practices to respect~\cite{githubdocs_copilot_instructions}. %For instance, one can instruct the agent on where to find relevant files, how to run tests, and what best practices to follow. 
While a prior study~\cite{santos2025decodingconfigurationaicoding} investigated the structure and content of Claude instruction files, no prior work studied the relationship between the instruction files and the performance of agentic-PRs. 

The goal of this work is to empirically investigate whether the addition of instruction files is associated with an improvement in the way agents generate pull requests. This improvement is measured, at the project level, by comparing different agentic-PR metrics before and after the creation of the instruction file(s). These metrics correspond to the merge rate as a proxy for pull request success (RQ1), the number of commits, the size of the changes, and the number of modified files for the complexity of tasks agents are capable of undertaking (RQ2), and the amount of discussions and the time to merge as a proxy for the efforts required to merge agentic-PRs (RQ3). To better understand the difference between projects that managed to increase their merge rates, we conduct an initial quantitative study on the verbosity of the instruction files (RQ4). Our empirical work is guided by these research questions.

%the relationship between the definition of instruction files and how agents perform on their created pull requests. Such performance is measured by the success of AI-agents in generating pull requests to be merged, the complexity of tasks that agents can perform after adding the instructions, and the efforts required to merge agentic pull requests. We quantify the relationship between the instruction files and these perspectives by comparing the pull request metrics before and after adding the instruction files. 

%In particular, we conduct a quantitative comparison of several PR-level metrics (e.g., merge rate) before and after adopting instruction files. We further examine how the instruction files differ between projects that show improvements in merge rate ($\ge 10\%$) and those that observe declines ($\le -10\%$). Thus, we formulate the following research questions:

\noindent\textbf{RQ1.} \textit{\rqi} % This RQ investigates the merge rate of projects before and after adding instruction files. 
27.7\% of the projects increased their merge rate by at least 20\%, while 26.35\% showed a decreasing trend.

\noindent\textbf{RQ2.} \textit{\rqii} %This RQ compare the AI-agentic PRs before and after adding the instruction files in terms of code churn, number of modified files, and number of commits. 
35.35\%, 10.10\% and 12.12\% of the projects statistically increase their description length, code churn and
number of commits, while 4.04\%, 7.07\%, and 7.07\% have the same metrics decreased.

\noindent\textbf{RQ3.} \textit{\rqiii} %This RQ performs the same comparison as RQ2 on the merge efforts measured by the amount of discussion and time to merge. 
13.13\% and 15.15\% of projects show a statistically significant increase in the time and discussion required to merge AI-agent PRs, respectively, whereas 8.8\% of the projects depict a decrease in both metrics.

\noindent\textbf{RQ4.} \textit{\rqiv} %This RQ examines the depth of instruction files among projects that show an increased merge rate vs. those with a decreasing trend. 
The projects that managed to increase their merge rate are substantially longer in words (a median of 976 vs. 569 words) and header counts, mainly H3, compared to the projects that experienced a decrease.

\textbf{Take-home message:} Our results shed light on the need for studies that guide developers in writing good instruction files. In fact, defining the instruction files does not always lead to better results, which highlights the importance of the quality of the content of those files. Future studies are required to define standards and quality metrics for the instruction files that developers can leverage to enhance these files in a way that improves efficiency and autonomy of AI-agents. In other words, our findings suggest considering the development of instructions as a software engineering activity, the concept that we define as \textbf{Instructions-as-Code}. 

We provide our scripts and data in the replication package~\cite{repliction-package-msr26}. 

\section{Study Design}

In this section, we discuss how we collect agentic-PRs alongside the instruction files of our studied projects.

\subsection{Collection of Agentic-PRs}

%\ali{We can refer to these line of prior work~\cite{10.1145/3551349.3556896, rebatchi2024dependabot} to justify the selection of toy project-related keywords. Both of these studies examined the description of subject project to identify engineering projects from those that are personal and toy. In our work, we start with a seed of keywords (e.g., academic) by looking at the first 10 projects, and iteratively refining the list of keywords based on the resulting projects.}

From the AIDev dataset~\cite{li2025riseaiteammatessoftware}, we select the projects with at least 20 stars. From these projects, we filter out toy ones. These are projects whose names contain the following keywords: tutorial, docs, documentation, lab, presentation, exercise, bootcamp, hackathon, assignment, practice, resume, template, talks, udemy, sample, notebook, cookbook, guide, note, course, example, educational, survey, dataset, academic, academia, academy, interview, lesson, my\_, my-, portfolio, leetcode, neetcode, challenge, insights, github.io, blog, masterclass, and website.
We identify a set of 4,680 projects. For these projects, we collect their recent pull requests and metadata that are not in the AIDev dataset using the GitHub GraphQL API. With the additional metadata, we further filter out projects that are templates, archived, forks, have fewer than 50 commits, or less than 50 PRs. We also leverage the project description (part of the new metadata) to exclude the remaining toy projects. We end with a final set of 2,449 projects and 48,944 pull requests.

Since the AIDev has data up to July 30th, 2025, we follow the same approach of Li et al.~\cite{li2025riseaiteammatessoftware} to collect newer Agentic-PRs up to December 10, 2025, along with their relevant metrics %. We further collect metrics 
(e.g., number of commits) %related to the pull requests, which are further 
discussed in the approach of each of our research questions. For example, we specify ``is:pr head:copilot/'' in the GitHub search query to retrieve PRs co-authored by GitHub Copilot, which are combined with those reported by Li et al.~\cite{li2025riseaiteammatessoftware}.

\subsection{Collection of Instruction Files}

\begin{table}[!h]
\centering
\footnotesize
\caption{Agent-specific predicates used to identify instruction files across projects.}\label{tab:agent-predicates}
\begin{tabular}{l p{0.75\linewidth}}
\toprule
\textbf{Agent} & \textbf{Predicates} \\
\midrule
\multirow{2}{*}{\textbf{Devin}} &
\texttt{**/PULL\_REQUEST\_TEMPLATE/DEVIN\_PR\_TEMPLATE.md} \\
& \texttt{**/PULL\_REQUEST\_TEMPLATE/devin\_pr\_template.md} \\
\hline
\textbf{Cursor} &
\texttt{.cursor/*}; \texttt{.cursorrules}; \texttt{**/*.mdc} \\
\hline
\multirow{2}{*}{\textbf{Copilot}} &
\texttt{.github/copilot-instructions.md}\\
& \texttt{.github/instructions/*} \\
\hline
\textbf{Claude} &
\texttt{.claude/*}; \texttt{CLAUDE.md}; \texttt{.github/workflows/claude*.yml} \\
\hline
\textbf{Codex} &
\texttt{**/AGENTS.override.md}; \texttt{**/TEAM\_GUIDE.md}; \texttt{**/.agents.md}\\
\hline
\textbf{Common} &
\texttt{**/AGENTS.md}\\
\bottomrule
\end{tabular}
\end{table}

Once PRs are collected, we retrieve the instruction files of the coding agents. In particular, we obtain the directory tree for each project via the GitHub API and search for files whose names match the criteria presented in Table~\ref{tab:agent-predicates}. These names are obtained from the documentation~\cite{DevinGitHubIntegrationDocs, CursorDocsRules, GitHubCopilotCustomInstructions, ClaudeCodeVSCodeDocs, OpenAI_AGENTSmd_Guide}. Our study relies on the creation date of instruction files, as we compare pull requests before and after the creation of the instruction files. Such a creation date is obtained via the first commit that added the instruction file.
Note that we keep all the instruction files if a project has multiple ones. With the instruction files collected, we filter out projects that do not have any instruction files. To have a robust comparison, we also keep only projects with at least five agentic-PRs before and five ones after the creation of the project's instruction file. %\med{for projects that have many instruction files, do you keep projects with 5 PRs before the first instruction file and 5 PRs after the same file or 5 after the last file. For example, if a project has the instruction file A then file B. Do you remove projects with less than 5 PRs after B even if there are mroe than 5 PRs after A?}\ali{I keep projects that have 5 PRs before the first inst file, and 5 PRs after the same file.}. 
\textbf{Our final dataset consists of 148 projects and 15,549 Agentic-PRs.}

\section{Results}

\subsection*{RQ1. \rqi}

\underline{Motivation:} 
The goal of this research question is to quantify the benefit of adding instruction files to the merge rate, as a measure of agents' success in generating accepted pull requests. % understand whether there is a relationship between the existence of instruction files and the merge rate. 
The intuition is that guided agents are more likely to perform tasks correctly that result in merged agentic-PRs.

%\ali{We use the merge metric as a measure of PR outcome, consistent with prior work~\cite{10.1145/2568225.2568315}.}

%\ali{A new recent software engineering benchmark by LinearB~\cite{software-engineering-benchmarks-report} reported that AI-assisted PRs have an acceptenace rate that is less than half of human PRs. We therefore want to use such a metric to determine the performance of AI-agents in merging PRs under instruction files.}

%\ali{\cite{LENARDUZZI2021110806, 9749844} inspected different factors the incluence the PR acceptetance. We then leverage the merge outcome of agentic-PRs to evaluate how well different agents perform.}

\noindent\underline{Approach:} To answer this RQ, we compare for each project the merge rate of agentic-PRs before and after the creation of its instruction files. Since projects can include multiple instruction files, we perform two comparisons: (1) before and after the creation of the first file and (2) before the creation of the first file and after the creation of all (one or multiple) the files.
For comparison, we use the following ratio: 
$ \left( \frac{\text{merge\_rate\_after} - \text{merge\_rate\_before}}{\text{merge\_rate\_before}} \right) \times 100 $.
A value below zero indicates a decrease in merge rate, a value above zero indicates an increase, and zero denotes no change.

\begin{figure}[!tbp]
  \centering
  \includegraphics[width=.62\linewidth]{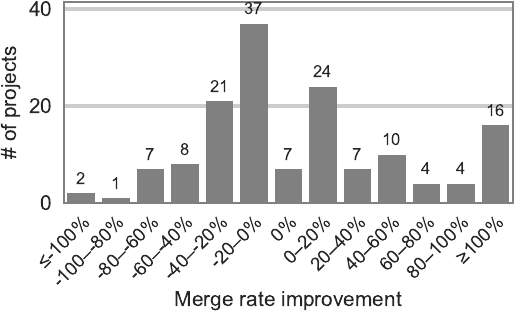}
  \caption{The \# of projects across merge rate improvement intervals for PRs closed before and after the first agent file creation.}\label{fig:barplot1}
\end{figure}

\noindent\underline{Results:} \textbf{Some projects benefit from the instruction files, exhibiting an increase in their merge rate, whereas other projects do not benefit and instead show a decreasing trend.}
We observe 27.7\% of the projects increased their merge rate by at least 20\%, as depicted in Figure~\ref{fig:barplot1}. 
For instance, the \textit{Significant-Gravitas/AutoGPT} project %(a platform to build and deploy AI-powered agents with five instruction files) 
showed a merge rate of 42.85\% before the first added agent file and 65.30\% after the last added agent file, resulting in a 52.38\% improvement. %This repository features five agent context files, namely \texttt{AGENTS.md}, which provides guidance for the Codex agent; \texttt{copilot-instructions.md}, which contains very detailed instructions to help Copilot better navigate the project (322 lines); and three Claude context files that are executed as CI/CD pipelines (CI failure auto-fix agent and Dependabot updates agent).
Furthermore, we observe that 26.35\% (39 out of 148) of the studied projects experienced a decrease in merge rate of at least 20\%. For example, the \textit{Azure/adx-mon} project %(an Azure observability platform) 
shows a merge rate of 75\% before the first agent file and 35.35\% after the same file was added. %This project includes a \texttt{copilot-instructions.md} file as its instruction file. This file appears to be less detailed than the one in the \textit{Significant-Gravitas/AutoGPT} project with 119 lines. The resulting decrease in merge rate is 39.39\%.
Only 4.72\% (7 out of 148) of the projects %show no change in merge rate. These projects typically 
exhibit the same merge rate before and after the first agent file.

\begin{figure}[!tbp]
  \centering
  \includegraphics[width=.65\linewidth]{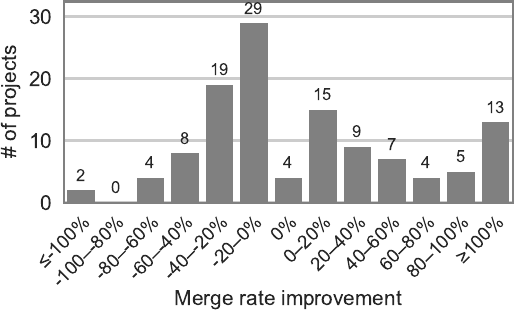}
  \caption{The \# of projects across merge rate improvement intervals for PRs closed before the first agent file and after the last agent file.}
  \label{fig:barplot2}
\end{figure}

\textbf{With all the instruction files in a given project, the percentage of projects that improve their merge rate by at least 20\% is 31.93\%}
%after adding all the instruction files}
, as shown 
Figure~\ref{fig:barplot2}. %highlights the distribution of projects whose merge rate is calculated before the firstly and lastly added agent files.
%Our findings reveal that 31.93\% experienced an increase in merge rate of at least 20\%.
For example, the \textit{NethermindEth/nethermind} repository (an Ethereum execution client) with two instruction files exhibits a merge rate increase of 246.15\% (i.e., a merge rate of 20\% before the first instruction file and 69.23\% after the addition of the second one).
%This repository includes two instruction files, \texttt{copilot-instructions.md} and \texttt{CLAUDE.md}, both of which are rich and detailed, with 251 and 255 lines, respectively.
Similarly, we observe that 27.73\% of the projects show a decrease in merge rate by at least 20\%.
For instance, the \textit{theopenco/llmgateway} project highlights a merge rate decrease %of 25.09\%, 
from 89\% %(before the first added agent file) 
to 66.66\%. % (after the last agent file).
%This project also contains two instruction files, \texttt{AGENTS.md} and \texttt{CLAUDE.md}, which are similarly rich and descriptive (168 and 206 lines, respectively).
We also observe four projects with no change in merge rate.

\subsection*{RQ2. \rqii}

\noindent\underline{Motivation:} The goal of this research question is to investigate whether guiding the models through the instruction files is associated with more complex tasks performed by AI-agents. % perform more complex tasks. 

%\med{do you consider only merged PRs here, right?}\ali{Only projects with merged PRs}

%\ali{We leverage the PR-level compelxity metrics, namely code churn, number of commits, and description length, following prior work~\cite{Yu2016} }

%\ali{The same report by LinearB leveraged the PR size (e.g., code churn) to measure the comlexity of generated code. Hence, we use the same metrics to evaluate whether the definition instrcution files are associated with increased or decreased complexity when competing various tasks.}

\noindent\underline{Approach:} For each project, we statistically (Mann-Whitney U test; $\alpha$ = 0.05 and Cliff's delta) compare agentic pull requests before and after the creation of the instructions file in terms of the quantity of produced code, measured using the following metrics: description length, code churn, number of files, and number of commits in agentic-PRs. Our comparison considers before and after the creation of the first instructions file, and before the creation of the first file and the creation of the last one. We report on the number of projects that statistically improved or reduced the complexity of their changes. Since we consider successfully developing complex code, we focus only on merged pull requests in this RQ. %\med{any details about the merged PRs go here}\ali{
We identify 99 projects with at least five merged PRs both before and after the creation of the first agent file. Similarly, we identify 80 projects that meet the same criterion after the creation of all (one or more) their instruction files.%}We statistically compare for each project the distribution of each of those metrics before and after the creation of the instruction files. %Our comparison for each studied metric and project is as follows: 

\begin{figure}[!h]
  \centering
  \begin{subfigure}[b]{0.23\textwidth}
      \centering
      \includegraphics[width=\textwidth]{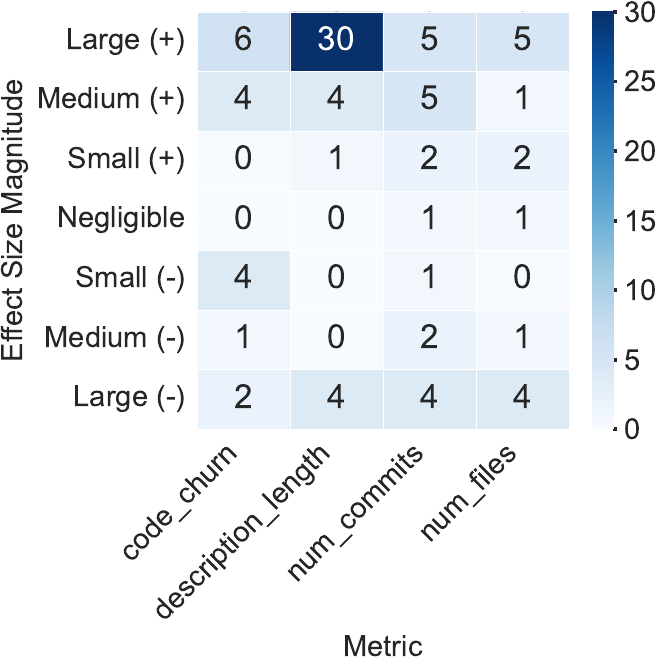}
      \caption{Before and after the first agent file.}\label{fig:heatmap1}
  \end{subfigure}
  \hfill
  \begin{subfigure}[b]{0.23\textwidth}
      \centering
      \includegraphics[width=\textwidth]{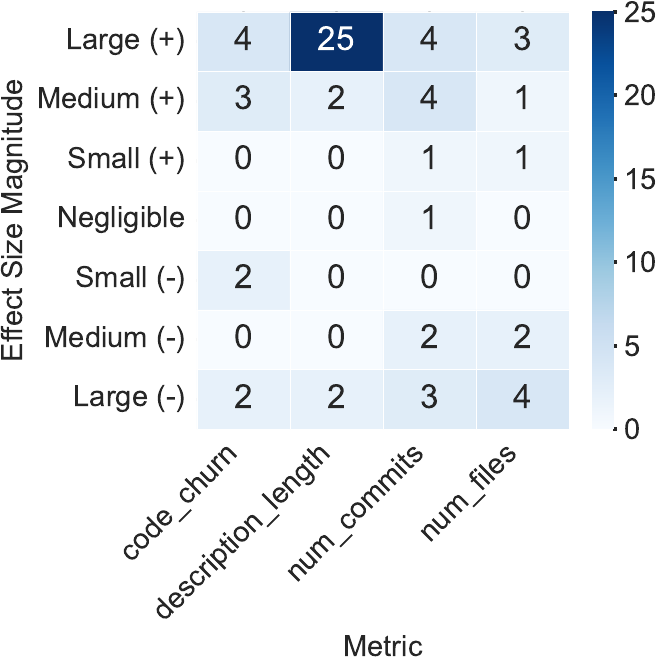}
      \caption{Before the first and after the last agent files.}\label{fig:heatmap4}
  \end{subfigure}
  \caption{The \# of stats test for agentic PR complexity metrics (a) when the first agent file is created and (b) when one of more agent files are created. $(+)$ indicates an increase and $(-)$ indicates a decrease.}
\end{figure}

\noindent\underline{Results:} \textbf{Projects that statistically increase AI-agent PR complexity metrics outnumber projects that decrease them}. %14.86\% and 13.51\% of the studied projects statistically increase their code churn and number of commits, respectively, when the first agent file is added.}
%When the first agent file is created, we observe that 
35.35\%, 10.10\%, 12.12\%, and 8.08\% of the projects show statistically significant increases in description length, code churn, number of commits, and number of changed files, respectively, with effect sizes ranging from small to large, as shown in Figure~\ref*{fig:heatmap1}. %This increase in description length can be attributed to the additional contextual information provided to the agent, which leads to more descriptive and fine-grained PR content. 
For instance, the \textit{asgardeo/thunder} project shows an increase in code churn of AI-agent PRs from a median of 308 to 2931.5, with a large effect size. %This project includes a lengthy \texttt{copilot-instructions.md} file with 214 lines and a moderately detailed \texttt{AGENTS.md} file.
Furthermore, we observe that 4.04\%, 7.07\%, 7.07.08\%, and 5.05\% of the projects exhibit statistically significant decreases in the description length, code churn, number of commits, and number of changed files, and, respectively, with effect sizes ranging from small to large. For example, the \textit{promptfoo/promptfoo} project significantly decreases the description length from a median of 563 to 203 %, code churn from 197 to 11.5, the number of commits from 6.5 to 3.5, and the number of changed files from 5.5 to 2
after the introduction of the first agent file.%\med{I didn't review the comments, waiting for the new findings about the size of changes} This project includes a \texttt{copilot-instructions.md} file as a context file, which highlights release preparation guidelines.

\begin{comment}
\begin{figure}[!h]
  \centering
  \includegraphics[width=.5\linewidth]{effect_size_heatmap_mode_2.pdf}
  \caption{The \# of stats test for all agentic PR metrics between PRs made before the agent's first file and after the last agent's file.}\label{fig:heatmap2}
\end{figure}
\end{comment}

\textbf{Our results hold when considering all the instruction files,} as shown in %Projects that statistically increase AI-agent PR complexity metrics outnumber projects that decrease them, even when they create all instruction files.
Figure~\ref*{fig:heatmap3}. % shows the distribution of projects with statistically significant differences in complexity metrics when instruction files are not used versus when all instruction files are created, across different effect size magnitudes. In particular, 
5, 7, 27, and 9 projects exhibit a statistically significant increase in the number of changed files, code churn, description length, and number of commits. For the same metrics, we observe a decrease for 6, 4, 2, and 5 projects, respectively. 

%while 11 projects show a decrease.

%Similarly, 17 projects indicate a significant increase in , compared to 8 projects that significantly reduce it. Exceptionally, projects with a large significant increase in description length substantially outnumber those that exhibit a decrease, with 35 and 3 projects, respectively. Regarding the number of commits, we observe that 17 projects show an increase compared to 11 projects with a reduction, while two projects exhibit a negligible effect size.

%These findings highlight the the complexity of tasks that AI agent tend to perform when context files are not used vs. when one or more files are created. Notably, coding agents tend to produce lengthy PR descriptions more frequently than other metrics, namely code churn, number of changed files, and number of commits.

\subsection*{RQ3. \rqiii}

\noindent\underline{Motivation:} The purpose of this research question is to evaluate whether the addition of instruction files is associated with the amount of effort required (measured by the time to merge and the amount of discussions) to merge a pull request. % The effort that we measure through the time between the creation and the merge and the amount of discussions %\ali{\sout{and revisions} since we don't have any metric related to the number of PR revisions. Thee review metric won't help since a few PRs have reviews.} 
%before merging an Agentic-PR. 

%\ali{We employ the time to merge and number of exchanged messages in revwing agentic-PRs to charaterize the review effort, inspired by prior work~\cite{Huang2021}.}

%\ali{Similarly, LinearB~\cite{software-engineering-benchmarks-report} used the Review Time to estimate how much effort to put in until an agentic-PR gets merged. What do you think of renaming ``merge time'' to ``review time'' to keep is consistent with the benchmark report by LinearB?}

\noindent\underline{Approach:} We follow the same approach of RQ2 using the following metrics: number of comments and the time to merge pull requests. In this RQ, since we are interested in the effort to merge AI-agentic PRs, we focus only on merged PRs. %, right?}\ali{Correct}.

\begin{figure}[!t]
  \centering
  \begin{subfigure}[b]{0.22\textwidth}
      \centering
      \includegraphics[width=0.7\textwidth]{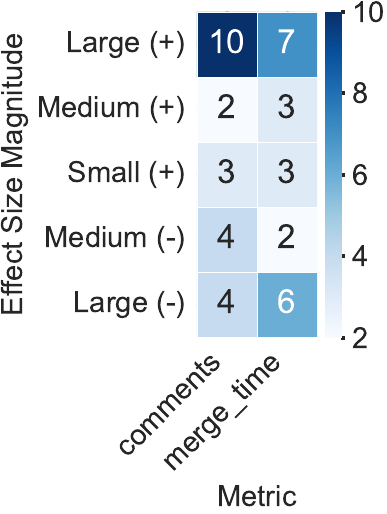}
      \caption{Before and after the first agent file.}\label{fig:heatmap3}
  \end{subfigure}
  \hfill
  \begin{subfigure}[b]{0.22\textwidth}
      \centering
      \includegraphics[width=0.7\textwidth]{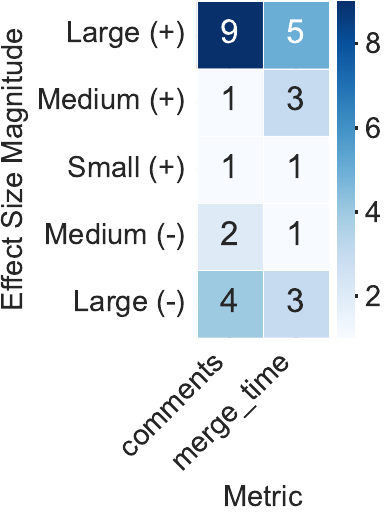}
      \caption{Before the first and after the last agent files.}\label{fig:heatmap4}
  \end{subfigure}
  \caption{The \# of stats test for agentic PR effort metrics (a) when the first agent file is created and (b) when one of more agent files are created.}
\end{figure}

\noindent\underline{Results:} \textbf{13.13\% and 15.15\% of projects show a statistically significant increase in the time and discussion required to merge AI-agent PRs, respectively, whereas 8.8\% of projects exhibit a decrease in both metrics.}
Figure~\ref*{fig:heatmap3} illustrates the number of projects that significantly increase or decrease the time to merge Agentic-PRs and the number of discussions involved when the first context file is added.
In particular, we observe that 15 projects show a statistically significant increase in the number of discussion comments involved in Agentic-PRs, among which 10 projects exhibit a large effect size, 3 a small effect size, and 2 a medium effect size. The \textit{quay/quay} project, with three agent files, shows an increase in the number of discussions from a median of 3 to 7 with a large effect size.
In contrast, 8 projects decrease the discussion size, with 4 projects showing a medium effect size and 4 a large effect size. For instance, the \textit{microsoft/PowerToys} project (a set of utilities to customize Windows) reduced the number of discussion comments from a median of 7 to 2 when the \texttt{copilot-instructions.md} file was created.
Furthermore, we observe that 13 projects significantly increase the time to merge agentic pull requests, while 8 projects decrease this time. For example, %the \textit{promptfoo/promptfoo}, which include 13 instruction files, increase the timer to merge from a median of 3.98 hours to 18.89 hours after when the first agent file is created. Conversely, 
the \textit{boxwise/boxtribute} project significantly reduced the time to merge from a median of 71.51 hours to 22.43 hours. The PR\footnote{\url{https://github.com/boxwise/boxtribute/pull/2292}} that introduced the \texttt{copilot-instructions.md} file shows a long discussion between Copilot and a developer to create detailed guidance for working within the Boxtribute codebase. This suggests that Copilot helped reduce the time to merge by leveraging such detailed instructions.

\textbf{11 projects statistically increase discussion size, while 8 projects increase merge time after the creation of all the instruction files.} As shown in Figure~\ref{fig:heatmap4}, we observe that 11 projects statistically increase the number of discussion comments, with 9 exhibiting a large effect size, while 6 projects decrease discussion size, among which 4 show a large effect size after the last agent file is created. Similarly, 9 projects highlight a statistically significant increase in merge time, compared to only 4 projects that show a decreasing trend.

\subsection*{RQ4. \rqiv}

\noindent\underline{Motivation:} While we observe that some projects improve after the creation of instruction files while others do not, this research question quantifies whether there is a difference between the two types of projects %projects that managed to improve compared to other projects 
in terms of the richness of the instruction files. %\ali{Such an analysis can help developers make more informed decisions when designing instruction files.}
%This RQ aims to better understand how the content characteristics of agent instruction files—specifically their size and organizational structure—relate to the effectiveness of agentic pull requests. Prior work by Chatlatanagulchai et al.~\cite{chatlatanagulchai2025agentreadmesempiricalstudy} examined various aspects of agent context files (e.g., \texttt{AGENTS.md}, \texttt{CLAUDE.md}, and \texttt{copilot-instructions.md}). Building on this line of work, we investigate whether the depth of instruction files, as reflected by their length and structural organization, is associated with increased or decreased merge rates of their pull requests.

\noindent\underline{Approach:} To explore the richness of the instruction files, we consider two metrics, which are the length of the instruction files at their creation time and the number of sections in the same files. % following prior work~\cite{chatlatanagulchai2025agentreadmesempiricalstudy, 10.1007/978-3-032-12089-2_40}.
%To address this RQ, we operationalize instruction file content using two complementary metrics. First, we measure the total number of words in each instruction file at the time of its creation, since these files may evolve over time. Second, we quantify the number of header sections present in the instruction file, which serves as 
These sections serve as a proxy for the degree of structural organization and the depth of agent guidance provided. We compare the length between projects that increased their merge rate by at least 20\% and the projects with at least 20\% merge-rate decrease after the creation of all the instruction files of their projects. %\med{are these sentences correct:} For projects with one file, we consider the increase or decrease before and after the creation of the same file. For projects with multiple files, we consider before the creation of the first file and after the creation of the last file. %\ali{I consider the merged rate before the first agent file and after the last agent file since we consider the richness of all instruction files.}  %We split the projects into two groups: (1) \textit{increase}: projects showing a increased merge rate of at least 20\% and (2) \textit{decrease}: projects with a decreased merge rate of at least 20\% after the lastly created agent file.

\begin{figure}[!t]
  \centering
  \begin{subfigure}[b]{0.22\textwidth}
      \centering
      \includegraphics[width=\textwidth]{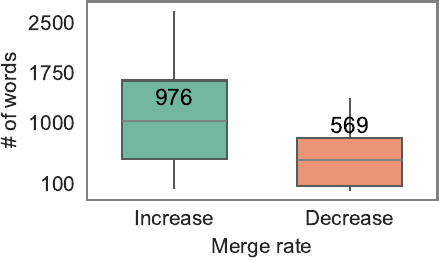}
      \caption{\# of words.}\label{fig:word_count}
  \end{subfigure}
  \hfill
  \begin{subfigure}[b]{0.22\textwidth}
      \centering
      \includegraphics[width=\textwidth]{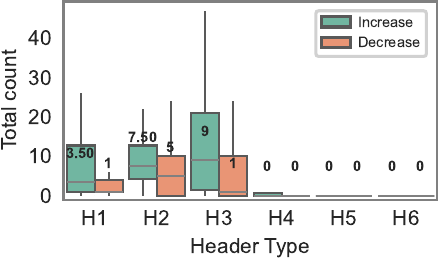}
      \caption{\# of headers}\label{fig:header_plot}
  \end{subfigure}
  \caption{Characteristics of instruction files for projects with $\ge20$ merge-rate increase or decrease, measured by (a) number of words and (b) number of headers.}
\end{figure}

\noindent\underline{Results:}
\textbf{Instruction files in projects with a $\ge20\%$ merge-rate increase are statistically longer than those in projects with a decreasing trend.} As shown in Figure~\ref{fig:word_count}, the projects whose merge rate increased by at least 20\% exhibit a median of 976 words, %in their agent context files—
which is nearly twice that of projects experiencing a merge-rate decrease of at least 20\%. Such an observation is supported by a Mann-Whitney U test, which yields a statistically significant p-value of 0.0029 with a small effect size. %For example, the \textit{Significant-Gravitas/AutoGPT} project, which shows a merge-rate increase of 68.52\%, includes two instruction files totaling 1,573 words in its instruction files. In contrast, the \textit{antiwork/gumboard} project (merge-rate increase of 25.6\%) contains only 63 words across its two instruction files, namely \texttt{copilot-instructions.md} and \texttt{.cursorrules}. Overall, these findings suggest that richer and more detailed instruction files are more likely to produce descriptive pull requests, which in turn increases the likelihood of merge. However, providing more extensive agent guidance may also lead to higher token usage, as noted in prior work~\cite{chatlatanagulchai2025agentreadmesempiricalstudy}.

\textbf{Projects with a $\ge20\%$ merge-rate increase exhibit substantially deeper structure in the first three header levels than projects with a decreasing trend.}
As depicted in Figure~\ref{fig:header_plot}, projects with a merge-rate increase of at least 20\% show higher median counts of header sections in their agent context files, with medians of 3.5, 7.5, and 9 for H1, H2, and H3 headers, respectively. In contrast, projects experiencing a merge-rate decrease of at least 20\% exhibit markedly fewer headers, with median counts of one H1, five H2, and one H3.
The Mann-Whitney U test produces a statistically significant difference for H1, H2, and H3 with a small effect size.
Moreover, projects with increasing merge rates tend to provide fine-grained guidance, as H3 headers are notably more prevalent compared to H1 and H2 sections. In line with this, Chatlatanagulchai et al.~\cite{chatlatanagulchai2025agentreadmesempiricalstudy} indicated that AGENTs.md, CLAUDE.md, and copilot-instructions.md are more dense around the H3 subsection. %Across both groups, higher-level headers beyond H3 are rarely used, with an observed median of zero, suggesting limited adoption of deeper hierarchical structuring in agent instruction files.

\section{Implications and Conclusion}

Our work empirically examines the relationship between the definition of instruction files and AI-agent pull requests. %We leverage the AIDev dataset with more recent data, which contains thousands of agentic pull requests. 
We find that while defining instructions is important for guiding the agents, not all the projects that add the instruction files improve their agents' efficiency. The efficiency that we measure using the merge rate, the efforts required to merge, and the ability of agents to generate a large amount of code. We find that projects with a better merge rate are those that have more elaborate instruction files, with files better structured into different sections and subsections. We recommend that developers make their instruction files more elaborate and well structured and improve them to fully benefit from agentic-AI models. For researchers, we recommend further studies on how to write instruction files, which we believe should be part of the software engineering processes. Inspired by Infrastructure as code, we suggest researchers develop standards, quality metrics, and elaborate processes around the development of instruction files, which should be considered as code (aka., \textbf{Instructions-as-code}). %Based on this analysis, we summarize the key implications observed in projects that define instruction files:

\bibliographystyle{ACM-Reference-Format}
\bibliography{sample-base}

@article{10.1145/3695988,
author = {Hou, Xinyi and Zhao, Yanjie and Liu, Yue and Yang, Zhou and Wang, Kailong and Li, Li and Luo, Xiapu and Lo, David and Grundy, John and Wang, Haoyu},
title = {Large Language Models for Software Engineering: A Systematic Literature Review},
year = {2024},
journal = {ACM Trans. Softw. Eng. Methodol.}
}

@article{10.1016/j.jss.2023.111734,
author = {Moradi Dakhel, Arghavan and Majdinasab, Vahid and Nikanjam, Amin and Khomh, Foutse and Desmarais, Michel C. and Jiang, Zhen Ming (Jack)},
title = {GitHub Copilot AI pair programmer: Asset or Liability?},
year = {2023},
volume = {203},
number = {C},
issn = {0164-1212},
journal = {J. Syst. Softw.},
month = 09,
numpages = {23}
}

@misc{cursor2024,
  title = {Cursor: The AI-Native Code Editor},
  author = {Anysphere},
  year = {2024},
  url = {https://cursor.sh/}
}

@techreport{anthropic2024claude,
  title={The Claude 3 Model Family: Tech Report},
  author={Anthropic},
  year={2024},
  url={https://shorturl.at/sdbBc}
}

@inproceedings{10.1145/3597503.3639219,
author = {Du, Xueying and Liu, Mingwei and Wang, Kaixin and Wang, Hanlin and Liu, Junwei and Chen, Yixuan and Feng, Jiayi and Sha, Chaofeng and Peng, Xin and Lou, Yiling},
title = {Evaluating Large Language Models in Class-Level Code Generation},
year = {2024},
series = {ICSE '24}
}

@inproceedings{10.1145/3597503.3639180,
author = {Liu, Zhe and Chen, Chunyang and Wang, Junjie and Chen, Mengzhuo and Wu, Boyu and Che, Xing and Wang, Dandan and Wang, Qing},
title = {Make LLM a Testing Expert: Bringing Human-like Interaction to Mobile GUI Testing via Functionality-aware Decisions},
year = {2024},
series = {ICSE '24}
}

@INPROCEEDINGS{11121707,
author={Cihan, Umut and Haratian, Vahid and İçöz, Arda and Gül, Mert Kaan and Devran, Ömercan and Bayendur, Emircan Furkan and Uçar, Baykal Mehmet and Tüzün, Eray},
booktitle={2025 IEEE/ACM 47th International Conference on Software Engineering: Software Engineering in Practice (ICSE-SEIP)}, 
title={Automated Code Review in Practice}, 
year={2025},
pages={425-436}
}

@online{githubdocs_copilot_instructions,
  author = {{GitHub}},
  title = {Adding repository custom instructions for GitHub Copilot},
  year = {2024},
  url = {https://shorturl.at/DRhK3},
  urldate = {2025-12-19},
  note = {GitHub Docs}
}

@misc{li2025riseaiteammatessoftware,
    title={The Rise of AI Teammates in Software Engineering (SE) 3.0: How Autonomous Coding Agents Are Reshaping Software Engineering}, 
    author={Hao Li and Haoxiang Zhang and Ahmed E. Hassan},
    year={2025},
    archivePrefix={arXiv},
    primaryClass={cs.SE}
}

@misc{santos2025decodingconfigurationaicoding,
    title={Decoding the Configuration of AI Coding Agents: Insights from Claude Code Projects}, 
    author={Helio Victor F. Santos and Vitor Costa and Joao Eduardo Montandon and Marco Tulio Valente},
    year={2025},
    archivePrefix={arXiv},
    primaryClass={cs.SE}
}

@misc{repliction-package-msr26,
  title        = {{Replication Package for: Toward Instructions-as-Code: Understanding the Impact of Instruction Files on Agentic Pull Requests}},
  month        = 12,
  year         = 2025,
  publisher    = {Figshare},
  version      = {v1.0.0},
  doi          = {10.6084/m9.figshare.30951143}
}

@misc{chatlatanagulchai2025agentreadmesempiricalstudy,
    title={Agent READMEs: An Empirical Study of Context Files for Agentic Coding}, 
    author={Worawalan Chatlatanagulchai and Hao Li and Yutaro Kashiwa and Brittany Reid and Kundjanasith Thonglek and Pattara Leelaprute and Arnon Rungsawang and Bundit Manaskasemsak and Bram Adams and Ahmed E. Hassan and Hajimu Iida},
    year={2025},
    eprint={2511.12884},
    archivePrefix={arXiv},
    primaryClass={cs.SE}
}

@misc{DevinGitHubIntegrationDocs,
  author       = {{Devin.ai Documentation}},
  title        = {GitHub Integration — Devin Docs},
  howpublished = {\url{https://docs.devin.ai/integrations/gh\#search-\&-precedence-order}},
  note         = {Accessed: 2025-12-25},
  year         = {2025},
  organization = {Devin.ai}
}

@misc{CursorDocsRules,
  author       = {{Cursor Documentation}},
  title        = {Rules — Cursor Docs},
  howpublished = {\url{https://cursor.com/docs/context/rules}},
  note         = {Accessed: 2025-12-25},
  year         = {2025},
  organization = {Cursor}
}

@misc{GitHubCopilotCustomInstructions,
  author       = {{GitHub Documentation}},
  title        = {Adding Repository Custom Instructions for GitHub Copilot},
  howpublished = {\url{https://docs.github.com/en/copilot/how-tos/configure-custom-instructions/add-repository-instructions}},
  note         = {Accessed: 2025-12-25},
  year         = {2025},
  organization = {GitHub, Inc.}
}

@misc{ClaudeCodeVSCodeDocs,
  author       = {{Anthropic / Claude Code Documentation}},
  title        = {Use Claude Code in VS Code},
  howpublished = {\url{https://code.claude.com/docs/en/vs-code}},
  note         = {Accessed: 2025-12-25},
  year         = {2025},
  organization = {Claude / Anthropic}
}

@misc{OpenAI_AGENTSmd_Guide,
  author       = {{OpenAI Developers}},
  title        = {Custom Instructions with AGENTS.md},
  howpublished = {\url{https://developers.openai.com/codex/guides/agents-md/}},
  note         = {Accessed: 2025-12-25},
  year         = {2025},
  organization = {OpenAI}
}

% %%
% %% If your work has an appendix, this is the place to put it.
% \appendix

% \section{Research Methods}

% \subsection{Part One}

% Lorem ipsum dolor sit amet, consectetur adipiscing elit. Morbi
% malesuada, quam in pulvinar varius, metus nunc fermentum urna, id
% sollicitudin purus odio sit amet enim. Aliquam ullamcorper eu ipsum
% vel mollis. Curabitur quis dictum nisl. Phasellus vel semper risus, et
% lacinia dolor. Integer ultricies commodo sem nec semper.

% \subsection{Part Two}

% Etiam commodo feugiat nisl pulvinar pellentesque. Etiam auctor sodales
% ligula, non varius nibh pulvinar semper. Suspendisse nec lectus non
% ipsum convallis congue hendrerit vitae sapien. Donec at laoreet
% eros. Vivamus non purus placerat, scelerisque diam eu, cursus
% ante. Etiam aliquam tortor auctor efficitur mattis.

\end{document}